\documentclass{caosp310}
\usepackage{graphicx}

\usepackage{natbib}
\articleNo{361}
\pubyear{2024}
\volume{54}
\volnumber{3}
\firstpage{1}
\received{ }
\accepted{ }

\def\BibTeX{{\rm B\kern-.05em{\sc i\kern-.025em b}\kern-.08em
             T\kern-.1667em\lower.7ex\hbox{E}\kern-.125emX}}
\def\hb{H$\beta$\/}
\def\feiiq{Fe{\sc ii}$\lambda$4570}

\def\civ{{C}{\sc iv}$\lambda$1549\/}
\def\mgii{{Mg}{\sc ii}$\lambda$2800\/}
\def\mgiionly{{Mg}{\sc ii}\/}
\def\aliiionly{{Al}{\sc iii}\/}
\def\civonly{{C}{\sc iv}\/}
\def\siiv{{Si}{\sc iv}$\lambda$1397\/}
\def\oiv{{O}{\sc iv}]$\lambda$1402\/}

\def\aliii{{Al}{\sc iii}$\lambda$1860\/}

\def\feii{{Fe}{\sc ii}\/}
\def\kms{km\,s$^{-1}$\/}

\def\rfe{$R_\mathrm{{Fe}{\textsc{ii}}}$\/}
\def\mbh{$M_\mathrm{\rm BH}$\/}
\def\lledd{$L/L_\mathrm{\rm Edd}$\/}

\def\civonly{{C}{\sc iv}\/}

\def\c14{$c$(1/4)\/}
\def\ergs{erg s$^{-1}$\/}
\def\apjs{ApJS}
\def\aap{A\&Ap}
\def\mnras{MNRAS\/}
\def\apjl{ApJL\/}
\def\apj{ApJ\/}

\def\aj{AJ\/}
\def\nat{Nature\/}
\def\apss{ApSS\/}

\begin{document}

%%%%%%%%%%%%%%%%%%%%%%%%%%%%%%%%%%%%%%%%%%%%%%%%%%%%%%%%%%%%%%%%%%%%%%%%%%%%%
%              R U N N I N G   P A G E   H E A D I N G S                     
% Odd page headings (except for the title page) are produced automatically
% and contain the title. If, and only if, the title of your article is too
% long the running head is omitted in the printout; you can make your own
% running title by using the \htitle command, putting the shortened title
% between the curly brackets. \htitle should also be used when the
% subtitle is present: \htitle offers you a way how to include it into the
% headings. If you wish to see how it works simply remove the % sign from
% the beginning of that line.
%
% Unlike the \htitle command, the \hauthor command is compulsory. It is
% used to produce even page headings and contains the names of the authors
% of an article.  All authors must be listed here, if possible. When
% authors' list is too long, you can abbreviate it by using "{\it et
% al.}". Authors' names are given in the form: initial(s) of the author's
% first name and surname. Authors are separated by a "," (comma) sign and
% the last one by "and".
%%%%%%%%%%%%%%%%%%%%%%%%%%%%%%%%%%%%%%%%%%%%%%%%%%%%%%%%%%%%%%%%%%%%%%%%%%%%%
%\htitle{A note to comet ejection process ...}
\hauthor{P. Marziani, E. Bon, S. Panda, N. Bon, et al.}
%\hauthor{L.\,Neslu\v{s}an {\it et al.}}

%%%%%%%%%%%%%%%%%%%%%%%%%%%%%%%%%%%%%%%%%%%%%%%%%%%%%%%%%%%%%%%%%%%%%%%%%%%%%
%                       T I T L E                                            
% Capital letters in the title are only used at the beginning of the
% names. Don`t end the title by a "." (dot)
%%%%%%%%%%%%%%%%%%%%%%%%%%%%%%%%%%%%%%%%%%%%%%%%%%%%%%%%%%%%%%%%%%%%%%%%%%%%%
\title{Evidence for a Stratified Accretion Disk Wind in AGN}

%%%%%%%%%%%%%%%%%%%%%%%%%%%%%%%%%%%%%%%%%%%%%%%%%%%%%%%%%%%%%%%%%%%%%%%%%%%%%
%                       S U B T I T L E                                      
% You can use the subtitle, with the command \subtitle similar to the
% \title command.
%%%%%%%%%%%%%%%%%%%%%%%%%%%%%%%%%%%%%%%%%%%%%%%%%%%%%%%%%%%%%%%%%%%%%%%%%%%%%

%%%%%%%%%%%%%%%%%%%%%%%%%%%%%%%%%%%%%%%%%%%%%%%%%%%%%%%%%%%%%%%%%%%%%%%%%%%%%
%                   A U T H O R  N A M E S                                   
% Authors' names are separated by the \and command and their institutes
% are assigned by the \inst{n} command. 
% If all authors belong to just one institute, it is not needed/desired
% to use the \inst command.
%
% Author can indicate her/his ORCID (https://orcid.org/) identifier using
% the command \orcid. It will not appear in the LaTeX output but will be
% sent to the ADS database.
%
% When the name contains "Slovak" letters L,d,t,l with a caron, use an
% a new \softl, etc. command (examples given in the last table of
% this document) to produce typographically correct accented characters.
%%%%%%%%%%%%%%%%%%%%%%%%%%%%%%%%%%%%%%%%%%%%%%%%%%%%%%%%%%%%%%%%%%%%%%%%%%%%%
\author{
        P.\,Marziani\inst{1}\orcid{0000-0002-6058-4912}
      \and
        E.\,Bon\inst{2} 
      \and 
        S.\,Panda\inst{3}   
      \and 
        N.\,Bon\inst{2}
       \and 
       A.\, Del Olmo\inst{4}
       \and
       A.\, Deconto-Machado\inst{4,5}
       \and 
       K. Garnica\inst{6}
       \and
       D. Dultzin\inst{6}
       %\and
       %W.\, Else\inst{?} 
       }

%%%%%%%%%%%%%%%%%%%%%%%%%%%%%%%%%%%%%%%%%%%%%%%%%%%%%%%%%%%%%%%%%%%%%%%%%%%%%
%           I N S T I T U T E S'  A D D R E S S E S                          
% The affiliation of authors is generated by the \institute command, the
% \and command being again used to separate individual addresses.
% The following commands may be used for the following three institutes:   
%               \lomnica        for      AsU SAV, Tatranska Lomnica          
%               \blava          for      AsU SAV, Bratislava                 
%               \ondrejov       for      AsU CAV, Ondrejov                   
%
% The given postal address must be complete in order to facilitate our
% editorial work. Moreover, you can add your e-mail address, using the
% \email command.
%%%%%%%%%%%%%%%%%%%%%%%%%%%%%%%%%%%%%%%%%%%%%%%%%%%%%%%%%%%%%%%%%%%%%%%%%%%%%
\institute{
           INAF, Astronomical Observatory of Padua, Italy \email{Paola.marziani@inaf.it}
         \and 
           Belgrade Astronomical Observatory, Serbia
         \and 
          International Gemini Observatory, NSF NOIRLab, La Serena, Chile 
         \and
           IAA-CSIC, Granada,  Spain
           \and
           IASF-INAF, Milan, Italy
           \and
           IA-UNAM, Mexico City, Mexico
          }

%%%%%%%%%%%%%%%%%%%%%%%%%%%%%%%%%%%%%%%%%%%%%%%%%%%%%%%%%%%%%%%%%%%%%%%%%%%%%
%                        D A T E / R E C E I V E D                          
% Date inserted here will be the date when your paper was received The
% format is: month (not abbreviated), day, year.
%%%%%%%%%%%%%%%%%%%%%%%%%%%%%%%%%%%%%%%%%%%%%%%%%%%%%%%%%%%%%%%%%%%%%%%%%%%%%
\date{March 8, 2003}
%\date{March 10, 2003}

%%%%%%%%%%%%%%%%%%%%%%%%%%%%%%%%%%%%%%%%%%%%%%%%%%%%%%%%%%%%%%%%%%%%%%%%%%%%%
%                        M A K E T I T L E
% The beginning part (title, author(s), etc.) of your article must be
% closed by the \maketitle command.
%%%%%%%%%%%%%%%%%%%%%%%%%%%%%%%%%%%%%%%%%%%%%%%%%%%%%%%%%%%%%%%%%%%%%%%%%%%%%
\maketitle

%%%%%%%%%%%%%%%%%%%%%%%%%%%%%%%%%%%%%%%%%%%%%%%%%%%%%%%%%%%%%%%%%%%%%%%%%%%%%
%                        A B S T R A C T,  K E Y W O R D S                   
% Here it is shown how to write an abstract.  Keywords should be placed
% within the "abstract" environment using the command \keywords and they
% should be selected from the thesaurus from Astron.  Astrophys.
% Abstracts. They must be separated from each other by -- (two dashes).
%%%%%%%%%%%%%%%%%%%%%%%%%%%%%%%%%%%%%%%%%%%%%%%%%%%%%%%%%%%%%%%%%%%%%%%%%%%%%
\begin{abstract}
We present observational evidence supporting the presence of a stratified accretion disk wind in active galactic nuclei (AGN), based on multi-wavelength spectroscopic analysis of broad and narrow emission lines. The diversity in emission line profiles, ionization potentials, and kinematic signatures suggests a structured outflow emerging from the accretion disk, with different zones contributing to specific spectral features. High-ionization lines (e.g., C{\sc iv} $\lambda$1549) exhibit strong blueshifts and asymmetric profiles indicative of fast, inner winds,  while low-ionization lines (e.g., H$\beta$, Mg{\sc ii} $\lambda$ 2800) show more symmetric profiles consistent with predominant emission from slower, denser regions farther out, although exhibiting systematic blueshifts in quasars radiating at high Eddington ratios. The intermediate ionization lines (e.g., Al{\sc iii}$\lambda$1860) present a situation that is intermediate in terms of shift amplitudes, although in several super-Eddington candidates radial outflow velocity may reach values comparable  to the ones of the high ionization lines. These results are consistent with radiatively   driven wind models featuring   radial stratification. We made preliminary photoionization modeling assuming unabsorbed radiation emitted from the corona and the hotter disk regions emission or absorbed by   a layer of gas. Our findings provide new constraints on the geometry and physical conditions of AGN winds, providing clear evidence in favor of stratified wind emission. 
\end{abstract}

%%%%%%%%%%%%%%%%%%%%%%%%%%%%%%%%%%%%%%%%%%%%%%%%%%%%%%%%%%%%%%%%%%%%%%%%%%%%%
%                       S E C T I O N I N G                                  
% Any section starts with the command \section as shown below, with the
% title in Initial Capitals and lowercase only. Do not number the sections
% - let LaTeX do that for you - and do not end them by a "." (dot).
%
% The (sub)section titles are typeset in boldface; so, if working in the
% mathematics mode in (sub)section titles, you must use \boldmath and 
% enclose it into curly brackets, e.g. "{\bolmath $R^{2}$}".
%%%%%%%%%%%%%%%%%%%%%%%%%%%%%%%%%%%%%%%%%%%%%%%%%%%%%%%%%%%%%%%%%%%%%%%%%%%%%
\section{Introduction}
%%%%%%%%%%%%%%%%%%%%%%%%%%%%%%%%%%%%%%%%%%%%%%%%%%%%%%%%%%%%%%%%%%%%%%%%%%%%%
%                       L A B E L                                            
% The label command is very convenient for you when referring to sections,
% subsections,..., tables, figures as well as to equations (see commands
% \ref and \pageref). In the case of figure and/or table environments the
% \label command should always be put after the \caption command to
% preserve proper numbering. When using the \label command the file must
% be compiled twice to get proper cross-references.
%%%%%%%%%%%%%%%%%%%%%%%%%%%%%%%%%%%%%%%%%%%%%%%%%%%%%%%%%%%%%%%%%%%%%%%%%%%%%

\label{intr}

The concept of a quasar “main sequence” (MS) emerged from the landmark work of \citet{borosongreen92}, who identified a strong anti-correlation between the relative strength of \feiiq\  emission and the full width at half maximum (FWHM) of the broad \hb\ line. This relation, commonly expressed through the parameter \rfe = \feiiq/\hb, was first hinted at in earlier works (e.g., \citealt{gaskell85}  and has since been confirmed and extended in a large number of studies 
\citep{sulenticetal00a,sulenticetal00b,shenho14,rakshitetal20,wushen22}. 
The MS framework provides a powerful tool for organizing the remarkable diversity of type-1 active galactic nuclei (AGN) spectroscopic properties.

Within this scheme, type-1 AGN can be broadly separated into two main populations: Population A and Population B \citep{sulenticetal00a,sulenticetal02,sulenticetal11}. Population A sources, characterized by narrower \hb\ profiles (FWHM $<$ 4000 \kms), are generally associated with high accretion rates relative to the Eddington limit, while Population B sources (FWHM $>$ 4000 \kms) are typically lower Eddington ratio (\lledd, where $L_\mathrm{Edd}$ is the Eddington luminosity) systems. A small subset, on the order of $\sim$ 10\% of the population, represents “extreme” Population A sources (\rfe $>$ 1), which are widely interpreted as candidates for super-Eddington accretion \citep{wangetal13, marzianisulentic14,duetal18,pandamarziani23,panda24,marzianietal25}.

The MS is thought to be primarily driven by the Eddington ratio \lledd\ \citep{marzianietal01,boroson02,shenho14,sunshen15, pandaetal19}, with orientation effects also playing a significant role in shaping the observed diversity. Black hole mass effects, concomitant with viewing angle effects, become relevant when AGN samples cover a wide range in luminosity \citep{marzianietal18a,naddafetal25}. The organization of type-1 AGN properties across the MS has been extensively studied at multiple wavelengths, most prominently within the so-called 4D Eigenvector 1 (4DE1) parameter space introduced by \citet{sulenticetal00c}. This framework combines optical, UV, and X-ray measures to capture the multidimensional diversity of quasars, while also providing a physical interpretation in terms of accretion physics and geometry (see \citealt{fraix-burnetetal17} for a summary).

Large surveys, such as the Sloan Digital Sky Survey (SDSS, \citealt{yorketal00}), have provided further statistical foundation for these studies. For instance, \citet{zamfiretal10} analyzed $\approx$470 quasars at $z < 0.7$ with average bolometric luminosities $\log L \sim 45.5$\ erg/s, confirming the prevalence of the MS trends and strengthening the case for \lledd\ as a primary driver. More recently, refinements in spectral analysis and improved databases have continued to reinforce this picture (e.g., \citealt{shenho14,wushen22}).

In summary, the quasar main sequence organizes type-1 AGN along a continuum of properties primarily governed by Eddington ratio and orientation. At one end of the sequence lie the extreme Population A sources, with the largest \feii\ strengths and the highest inferred \lledd\ values, plausibly representing systems accreting at or above the Eddington limit. These sources provide critical laboratories for understanding black hole growth under extreme conditions and the broader role of AGN feedback in galaxy evolution \citep{marzianietal25}. Luminosity and black hole mass (\mbh) effects appear in sample covering large ranges in luminosity and mass. They can be reconduced to two effects: an increase in the amplitude of shifts with respect to the rest frame observed mainly in high-ionization lines \citep[e.g.,][]{marzianietal16a}, and a prominent redward asymmetries associated with very massive black holes \citep{marzianietal09,marziani23}.

In this paper, we first present a brief summary of three recent works dealing with the observations  of outflows as diagnosed from the shifts with respect to the rest frame, namely the prototypical \civ\  high-ionization line, the intermediate ionization emission \aliii, and the low ionization line of \mgii\ (Sections \ref{civ}, \ref{almg}). The three lines are all unresolved doublets associated with the resonance transition $^{2}P_{\frac{1}{2},\frac{3}{2}} \rightarrow ^{2}S_{\frac{1}{2}}$, with parent ionic species of different ionization potential, from $\approx 50$ eV (\civonly) to   $\approx 25$ eV (\aliiionly) and $\approx 15$ eV (\mgiionly). We afterwards attempt to explain the observational results by photoionization  models  focused on the sectors of the MS where the larger amplitude blueshifts are found, namely Population A and extreme Population A (Section \ref{photoion}).   

\begin{figure}[t!]
\centerline{\includegraphics[width=0.75\textwidth,angle=-90]{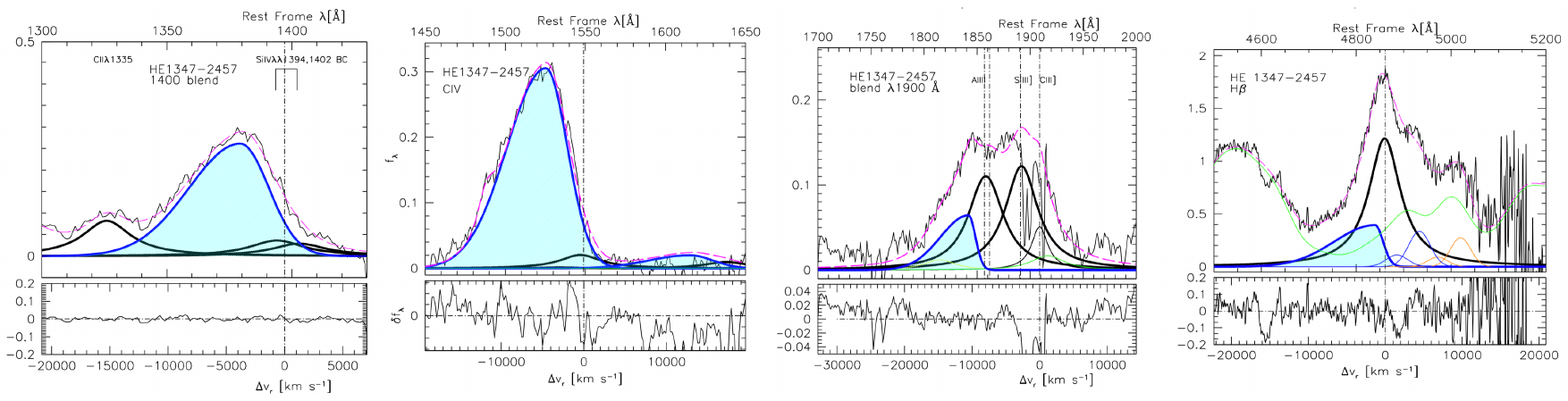}}\vspace{-2.75cm}
\caption{The continuum subtracted spectrum of the luminous Hamburg ESO Main Sequence (HEMS) quasar HE1347—2457, with an emphasis on the outflowing components of \siiv+\oiv,  \civ, \aliii, \hb\  (cyan-shaded components). }
\label{fig:outflows}
\end{figure}

\section{\civ\ Shifts and Outflows along the Quasar Main Sequence}
\label{civ}

Systematic blueshifts of the high-ionization \civ\ emission line provide one of the clearest observational signatures of quasar outflows. Early evidence for such features in composite spectra of radio-quiet quasars was strengthened by HST/FOS studies \citep{marzianietal96,corbinboroson96,sulenticetal07,marzianietal10,leighlymoore04,richardsetal11, sulenticetal07, marzianietal10,sulenticetal17}. These analyses showed that \civonly\  profiles can be decomposed into at least two components: a virialized, symmetric, and generally unshifted component (well represented by a Lorentzian in Population A sources), and a blue-shifted component that is naturally interpreted as emission from an outflowing wind. The latter is best modeled with a skewed Gaussian profile and becomes increasingly prominent at high \rfe\ values, i.e., in extreme Population A quasars.

Population trends are striking at low $z$: large blueshifts of $v_\mathrm{r} < –1000$\ \kms\ are observed almost exclusively in Population A sources, particularly those with FWHM(\hb) $< $ 4000 \kms\ and strong \feii\ emission (\rfe $>$ 1; \citealt{zamfiretal10,marzianietal10, richardsetal11}). In contrast, Population B quasars  rarely show strong \civonly\ blueshifts (if we exclude very high luminosity sources), reinforcing the link between high \lledd\ and the presence of radiatively driven winds.

At higher luminosities, large blueshifts persist. Near-infrared spectroscopy of 52 Hamburg–ESO quasars with $1 \la z \la 3$ and $\log L >$ 47 erg/s (HE main sequence, HEMS sample) confirmed that strong \civonly\  blueshifts are associated with high Eddington ratios rather than simply high black hole masses or luminosities \citep{marzianietal18,richardsetal11,giustiniproga19,deconto-machadoetal23,deconto-machadoetal24}. When viewed in the \lledd–$L$ and \lledd–\mbh\ planes, blueshift amplitude is most clearly organized by accretion state, with Population A objects (above \lledd $\approx$ 0.2) being wind-dominated \citep{marzianietal16a}.

There is, however, a trend between outflow velocity and luminosity \citep{marzianietal16a,sulenticetal17}. The scaling of outflow velocity with luminosity further supports a radiative-driving mechanism. Weak but consistent correlations ($v \propto L^{0.25}$) are in line with predictions from radiation-driven disk-wind models \citep{murrayetal95,laorbrandt02,proga07,proga07a}.  Pop. B at high luminosity also show large blueshifts, up to a few thousands \kms \citep{sulenticetal17}.  There is evidence of Pop. B outflows at  luminosity lower than $\log L \sim $ 47 erg/s (e.g., \citealt{richardsetal11,marzianietal22b}). Since Pop. B shifts at low luminosity remain of small amplitude and difficult to measure, the  considerations presented in this paper are meant for Pop. A and extreme Pop. A only. 

An example of powerful wind-dominated quasar of the HEMS survey is shown in Fig. \ref{fig:outflows}: the profiles of high ionization lines such as \civonly and \siiv+ \oiv\ are dominated by blueshifted emission associated with the outflows. In this rather extreme cases, the outflows also significantly  affect  \hb\ and the \aliii\ lines.  While the bolometric luminosity sets the available radiative power, it is the Eddington ratio that regulates the efficiency of wind launching and explains the prevalence of large blueshifts in high-\lledd\ sources.

Taken together,  the restrictions of large \civonly\ blueshifts to Population A and extreme Population A sources reinforces the interpretation of the quasar main sequence as primarily driven by \lledd, with \civonly\ outflows providing a direct tracer of radiative wind activity in the broad-line region.

\section{Intermediate- and Low-Ionization Lines: \aliii\ and \mgii}
\label{almg}
While the most dramatic blueshifts are observed in high-ionization lines such as \civonly, intermediate- and low-ionization species also show systematic kinematic signatures that can provide complementary insights into quasar outflows.

\subsection{\aliii}

The \aliii\ doublet, with intermediate ionization potential, generally shows only modest outflow signatures compared to \civonly. Statistical analyses reveal that \aliii\ blueshifts are correlated with those of \civonly\ but with a much shallower slope ($\approx$0.1), indicating a less prominent wind contribution \citep{marzianietal22, buendia-riosetal23}. Quantitatively, the median centroid shift of \aliii\ follows the relation
$c(\frac{1}{2})$(\aliiionly) $\approx$ (0.11 $\pm$ 0.03) $c(\frac{1}{2})$(\civonly) + (50 $\pm$ 70) \kms\ \citep{marzianietal22}, consistent with a weaker kinematic response of the intermediate-ionization gas. However, in the most extreme Population A quasars (\rfe $\ga$ 1), \aliiionly\ can display large blueshifts and asymmetric profiles comparable in strength to those of \civonly, suggesting that under super-Eddington conditions the wind dominates across a broader ionization range. %This reinforces the view that extreme Pop. A sources represent the locus of the strongest radiatively driven outflows in the quasar main sequence.

\subsection{\mgii}

The \mgii\ resonance doublet, one of the most widely used virial estimators of black hole mass \citep{marzianietal13a}, is typically far less affected by outflows \citep{trakhtenbrotnetzer12}. Most quasars show symmetric and only weakly shifted \mgiionly\ profiles. Nonetheless, careful spectral analysis has revealed subtle displacements of the line core, along with FWHM increases of a few hundred \kms, in the highest \lledd\ extreme Pop. A sources \citep{marzianietal13}. These modest but systematic shifts indicate that even low-ionization gas can participate in winds under conditions of extreme accretion. 

Following \citet{popovicetal19}, \mgii\ shows Lorentzian \mgiionly\ profiles and also signatures of outflow, but these are present only in part of the line and not in every source, as already noted by \citet{marzianietal13a}. \citet{popovicetal19} argue that \mgiionly\ consists of two kinematic pieces'': (i) a core that behaves similarly to \hb\ and appears virialized, and (ii) an additional “fountain-like’’ component, with motions roughly perpendicular to the disc, which produces the very broad wings and can reach shifts of a few thousand \kms. This second component may be associated with a photoionized ``bowl'' connecting the outer accretion disc and the inner torus \citep{goadetal12}, a configuration that is also consistent with a failed, radiatively accelerated, dusty wind \citep{czernyhryniewicz11,czernyetal17,naddafetal25}. Since the fountain component does not correlate with the virial broadening estimators, it is interpreted as due to outflows/inflows rather than rotation.

This configuration can explain the Lorentzian wings of \mgii. It is, however, not intended to account for the net blueshift of the \mgiionly\ line core reported by \citet{marzianietal13} for the highest \lledd\ sources. Those blueshifts are more naturally explained by a systematic line displacement, analogous to what is observed in \civ, although with a smaller amplitude.

\subsection{Ionization Potential and the Hierarchy of Shifts}

%The relative amplitudes of blueshifts scale, to first order, with the ionization potential. 
In extreme Pop. A quasars, $v$(\aliiionly) $\approx$ 0.3 $v$(\civonly) (and in some cases they are almost equal), while $v$(\mgiionly) $\approx$ 0.1 $v$(\civonly) \citep{marzianietal13,buendia-riosetal25}. The comparative behavior of \civ, \aliii, and \mgii\ therefore points to a simple trend: the higher the ionization potential of the line, the larger the blueshift. High-ionization lines such as \civonly\ (IP $\approx$ 47.9 eV) show the largest shifts, often $>-1000$ \kms\ in extreme Pop. A sources. Intermediate-ionization \aliiionly\ (IP $\approx$ 28.4 eV) usually shows smaller shifts, but in the most extreme Pop. A quasars its outflow signature can approach that of \civonly. In contrast, the low-ionization \mgiionly\ doublet (IP $\approx$ 15 eV) shows only modest shifts, typically $\sim$10\%\ of the \civ\ amplitude \citep{marzianietal13,marzianietal22,buendia-riosetal23,buendia-riosetal25}. This ionization-stratified sequence matches the expectations of disk–wind models, in which the highest-ionization gas traces the fastest parts of the flow, while lower-ionization lines form deeper in the broad-line region, where motions are more nearly virial.

This stratification follows naturally if the high-ionization gas occupies the more accelerated layers of a radiatively driven disk wind close to the continuum source, whereas lower-ionization species arise farther out in the BLR \citep{murrayetal95, proga07}. In this picture, extreme Population A quasars lie in the regime where radiation pressure not only controls the dynamics of the highest-ionization gas but also has a measurable impact on intermediate- and even low-ionization lines.
\section{Spectral Energy Distributions of High Accretors and Photoionization Simulations}
\label{photoion}

The interpretation of outflow signatures in high-accretion quasars requires a realistic description of their spectral energy distributions (SEDs). Recent works have established that quasars with high Eddington ratios (\lledd $\ga$ 1) and extreme \feii\ emission (\rfe $\ga$ 1) exhibit consistent and distinctive SEDs \citep{jinetal12,marzianisulentic14,ferlandetal20,pandamarziani23,garnicaetal25}. We utilized  the median SED derived from a sample of $\approx$150 low-redshift Pop. xA quasars \citep{garnicaetal25} as well as the high \lledd\ SED from \citet{jinetal12},  and we assume that the line emitting gas is exposed  either to the unobscured continuum or continuum absorbed by a hot, dense layer of gas at $r \approx   200 r_\mathrm{g}$\  (for  $\approx$  10$^8$ M$_{\odot}$; $\log N_{\mathrm{H}} = 10^{23}$ cm$^{-2}$; $\log n_{\mathrm H} = 10^{11}$ cm$^{-3}$).  A lower  black hole mass 10$^7$ M$_{\odot}$ appropriate for NLSy1s  would require a different, ``harder"  SED.

%Typical CLOUDY input parameters correspond to $\log N\_\mathrm{c} = 23$ cm$^{-2}$, $\log n\_H = 11$ cm$^{-3}$, and $\log r = 15.5$ cm.

\begin{figure}[t!]
%\centerline{\includegraphics[width=1.00\textwidth,clip=]{lratios.png}}
\includegraphics[width=0.35\textwidth]{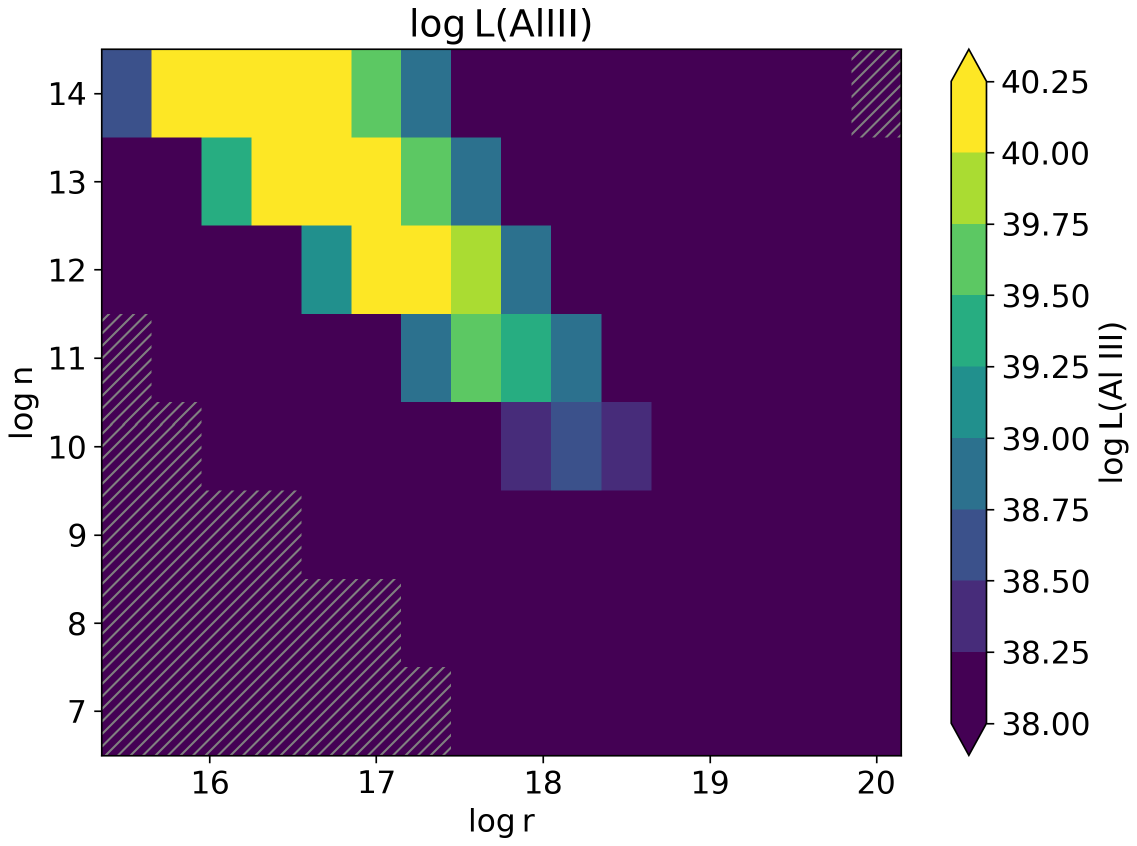}\hspace{-0.4cm}
\includegraphics[width=0.35\textwidth]{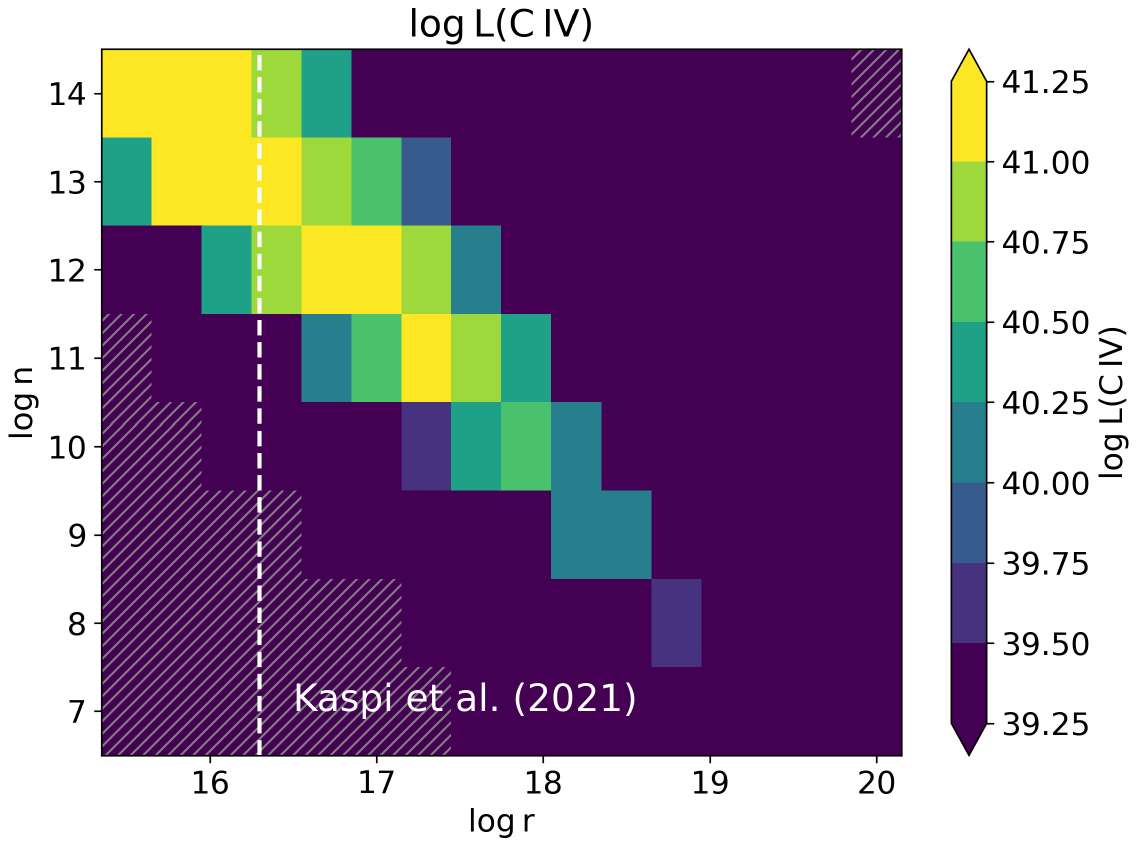}\hspace{-0.475cm}
\includegraphics[width=0.35\textwidth]{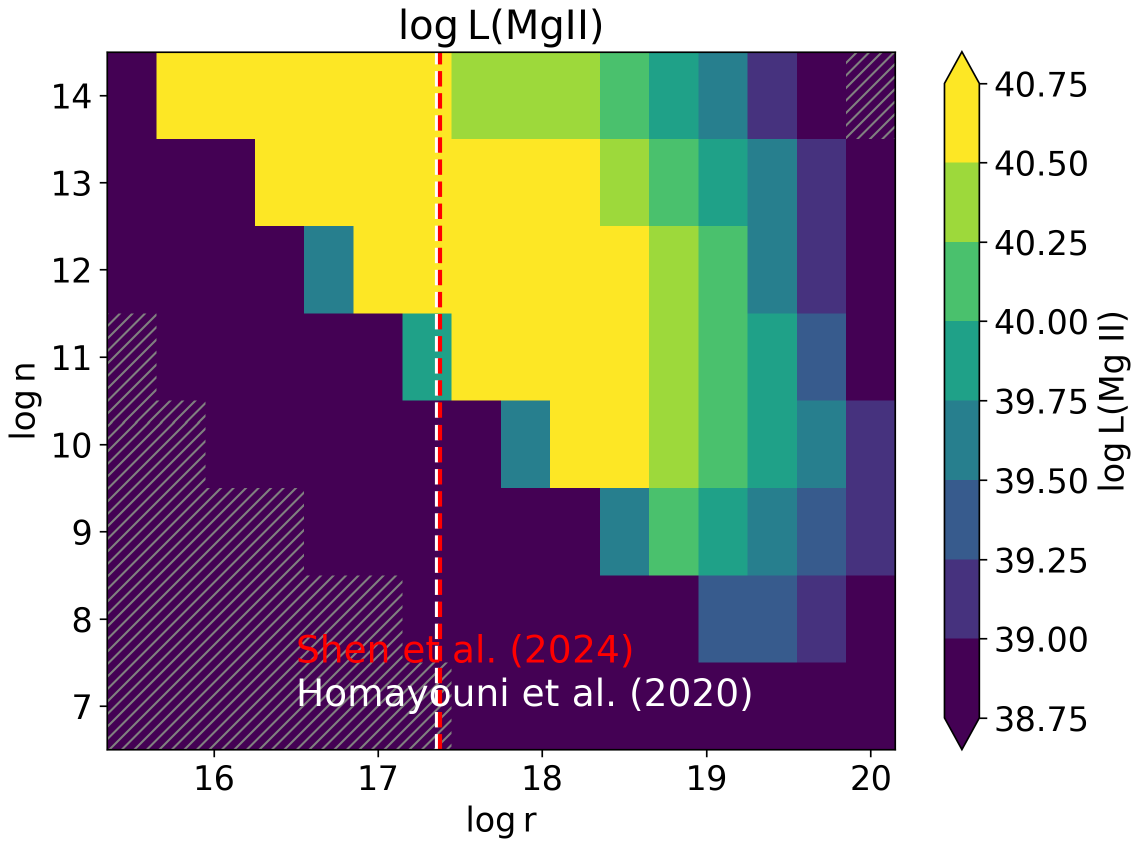}\vspace{-2.75cm}\\
\includegraphics[width=0.35\textwidth]{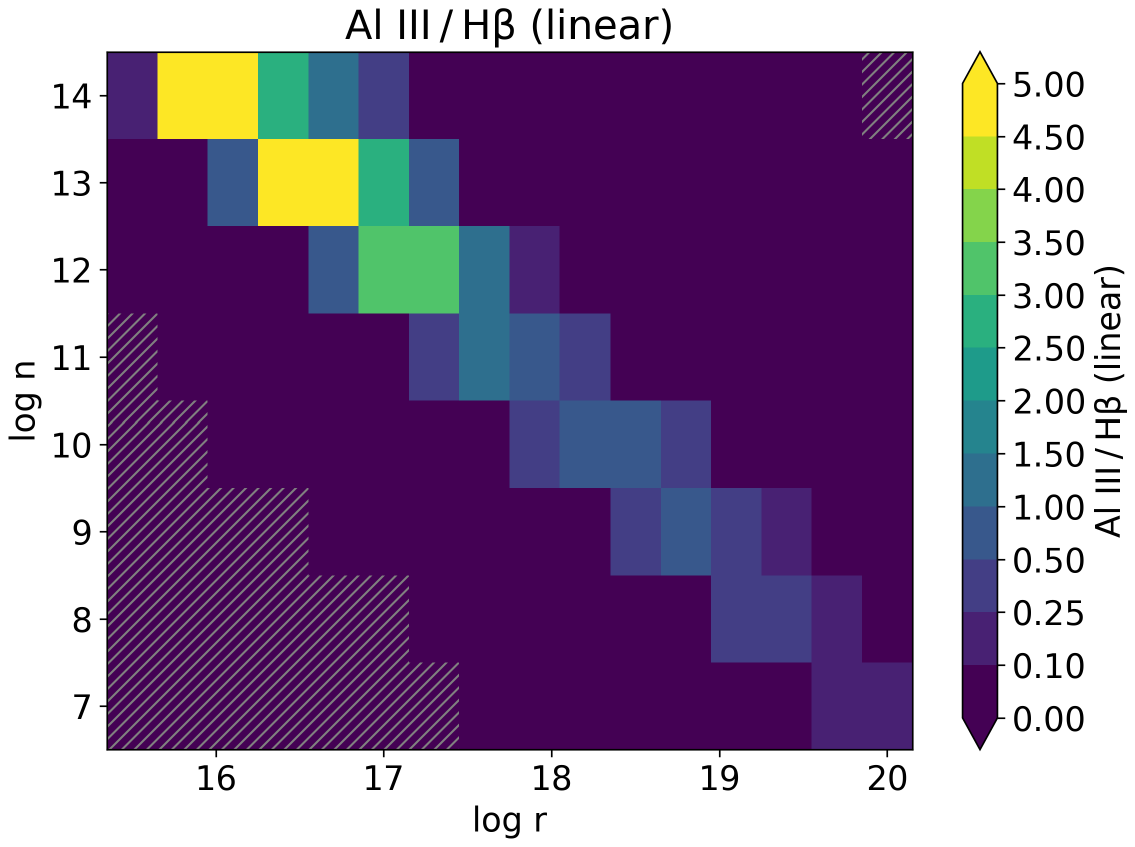}\hspace{-0.4cm}
\includegraphics[width=0.35\textwidth]{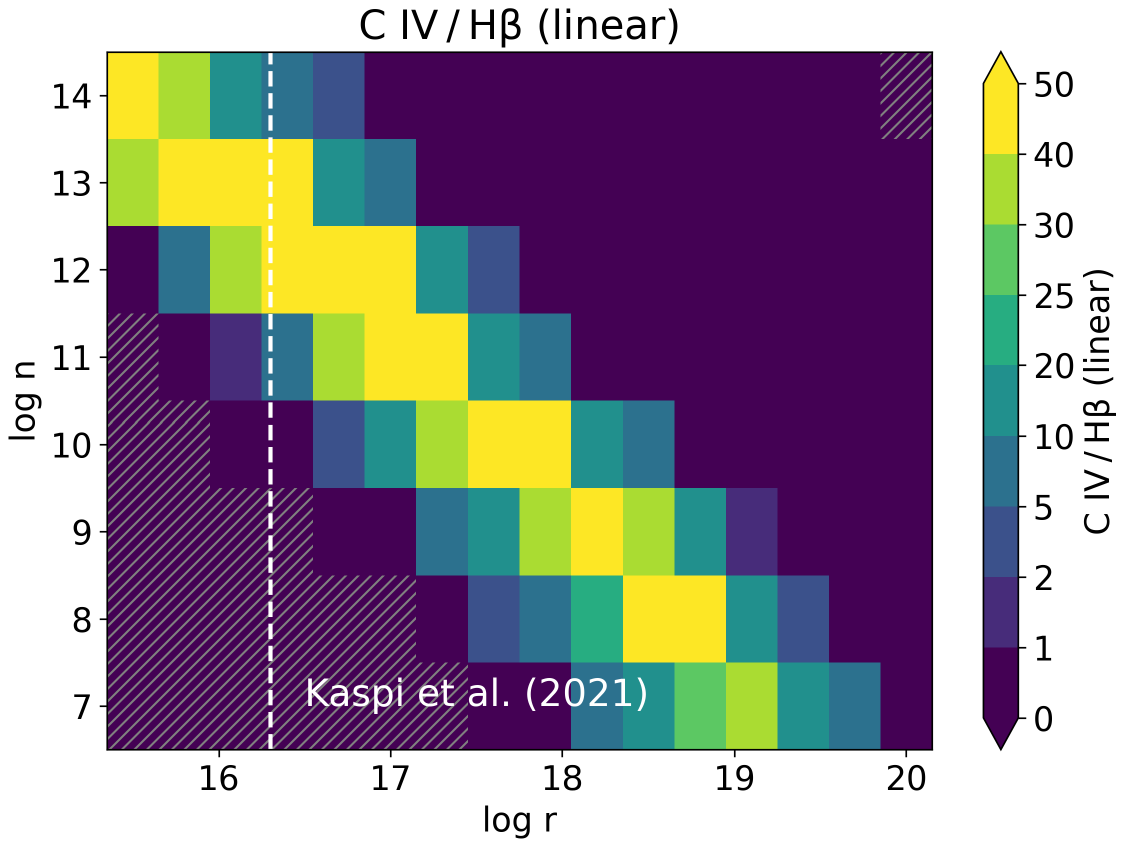}\hspace{-0.45cm}
\includegraphics[width=0.35\textwidth]{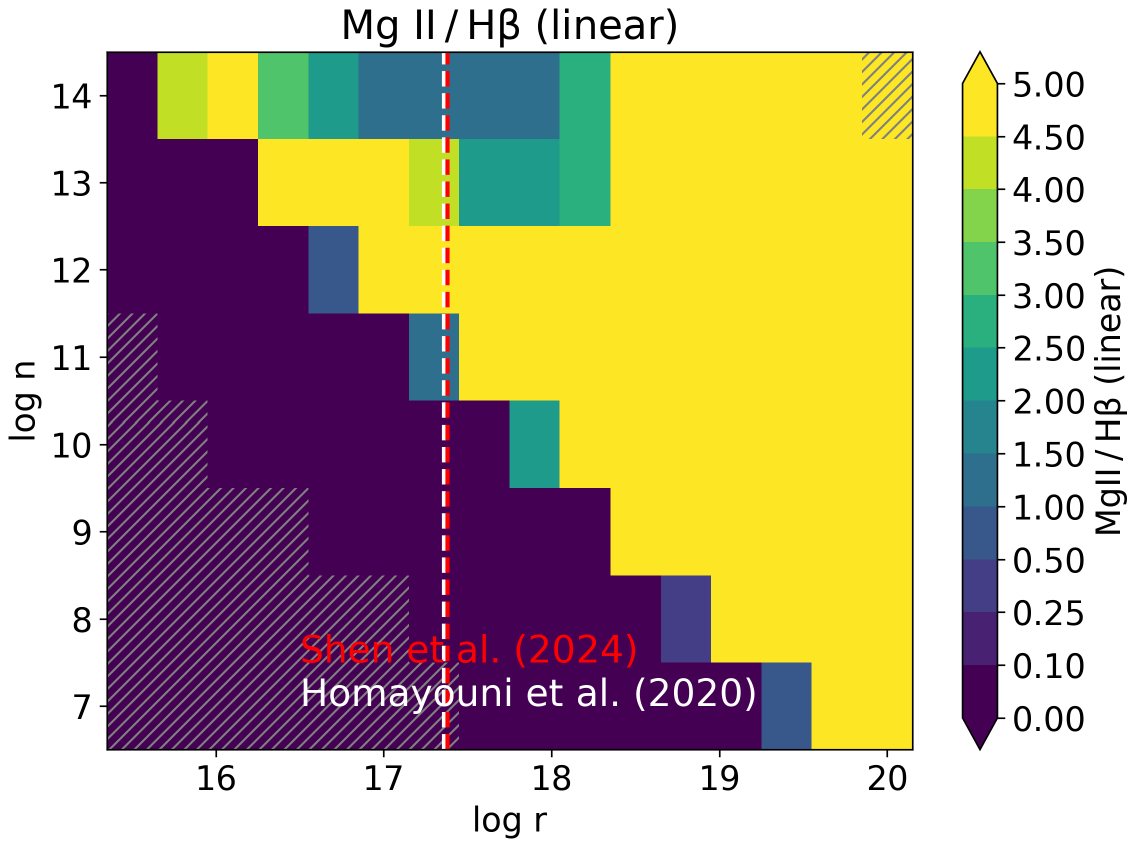}\vspace{-1cm} 
\caption{Behavior of line luminosity (top row)  and intensity ratios  with \hb\ (bottom), for \civ, \aliii, and \mgii\ as a function of hydrogen density and radius, for a $10^8$ solar masses black hole. The white vertical lines identify the radii measured according to the scaling laws of \citet{kaspietal21}  for \civ\ and  \citet{homayounietal20,shenetal24} for \mgii. }
\label{fig:lratios}
\end{figure}

\subsection{ Radiatively Driven Outflows}

Photoionization simulations highlight the conditions under which radiatively driven winds can be launched. Outflows are possible  at relatively low column densities ($N_\mathrm{c} \approx 10^{21}$ cm$^{-2}$), where radiative acceleration exceeds gravity. The requirement that the force multiplier $M > 2$ translates into the condition $M$ (\lledd) $> 1$\ \citep[see also][]{ferlandetal09,netzermarziani10}. At higher column densities ($N_\mathrm{c} \approx 10^{24}$ cm$^{-2}$; \citealt{ferlandpersson89,pandaetal20,pandaetal21}), radiation fails to overcome gravity, and the gas remains bound. These results confirm that outflow efficiency is intimately tied to the accretion state and gas structure in the inner broad-line region. 

\subsection{ Predicted Emission Luminosities and Line Ratios}

The simulations further predict the radial stratification of outflow emission. \civ\ emission is favored at smaller radii, \aliii\ at intermediate radii, and \mgiionly\ at larger distances, reflecting both ionization potential and gas density conditions. The  luminosities ($L$(\civonly), $L$(\aliiionly), $L$(\mgiionly)) define a narrow “corridor” of optimal ionization parameter where strong emission can be sustained (Fig. \ref{fig:lratios}). The continuum luminosity at $\lambda = 3000$ \AA\ has been assumed to be $\lambda L_{\lambda} (3000$\AA)$\approx 4.2 \cdot 10^{44}$ erg/s.  The hatched parts identify areas of the parameter plane where radiation forces are insufficient to drive an outflow; dark blue parts are regions where line emission is exceedingly low. The case shown here refers to a partially-absorbed high \lledd\ \citet{jinetal12} continuum, with the absorber located at  $r \approx   200 r_\mathrm{g}$. Due to the relative proximity of the absorber relative to the central continuum source, the absorbing gas remains hot and producing significant absorption only in the range between  1 and 10 keV. A fully unabsorbed \citet{jinetal12} continuum would give rise to similar trends. 

\begin{figure}[t!]
\centerline{\includegraphics[width=0.72\textwidth,angle=-90]{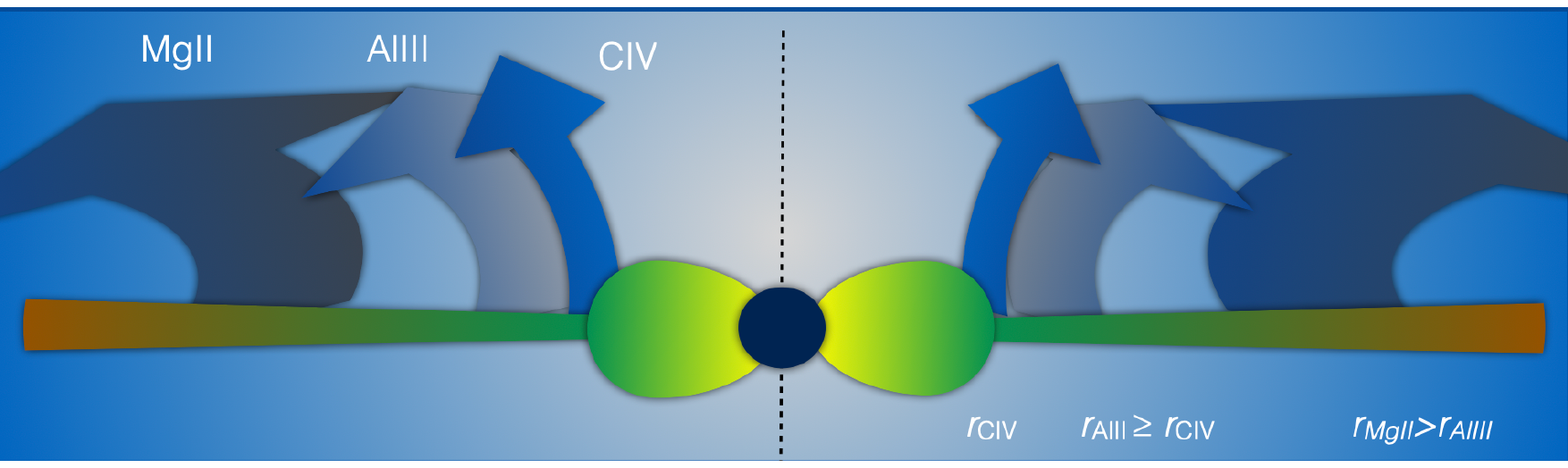}}\vspace{-2.25cm}
\caption{Sketch illustrating the differences in launching radii for the radiatively driven-wind emitting \civ, \aliii, \mgii. The accretion disk around the central black hole is assumed to have an inner, puffed-up and optically thick region sustained by radiation pressure, as well as an outer thin, optically thick region. The various elements are not drawn to scale.}
\label{fig:disks}
\end{figure}

The corridor peak, if considered at fixed density, is displaced toward larger radii for \aliiionly, and even more so for \mgiionly\ with respect to \civonly\ (top row of Fig. \ref{fig:lratios}), as schematically emphasized in Fig. \ref{fig:disks}.  Outside the corridor for \civonly\ and \aliiionly\ emission, for radii smaller than the ones of the corridor,  over-ionization suppresses line production, a behavior consistent with the observed weakness of \civ\ in extreme Pop. A quasars \citep{kaspietal21}.  For larger radii, under-ionization depresses the emission of both \civ\ and \aliii.   The behavior of \mgii\ is different: strong emission and large \mgii/\hb\ radii are possible over a wide range of radii that are however larger than the ones where \aliii\ and \civ\ are maximized.

Predicted line ratios also provide critical tests (bottom panels of Fig. \ref{fig:lratios}). For instance, {\tt CLOUDY} simulations reproduce observed \civ/\hb\ ratios $>$ 10–20 only under specific combinations of density and ionization, while \mgii/\hb\ ratios match those measured in large quasar samples \citep{homayounietal22, princeetal23}. The photoionization model therefore not only explains the ionization-dependent hierarchy of blueshifts (\civ\ $>$ \aliii\ $>$ \mgii) but also constrains the physical conditions under which these outflows form.

\section{ Conclusions}

Our analysis of emission-line shifts along the quasar main sequence reinforces the view that radiatively driven outflows are a ubiquitous feature of type-1 AGN. Outflow signatures are present even in single-epoch spectra, and they become dominant in high-ionization lines at high Eddington ratios and, more generally, at the highest luminosities ($L \ga  10^{47}$ \ergs).

In terms of accretion mode, Pop. A AGN satisfying the criterion \lledd $> 0.1 - 0.2$\ could be defined as black hole with an inner optically thick, geometrically thick region \citep[c.f.,][]{giustiniproga19}. Among Population A sources, the amplitude of line blueshifts decreases systematically with ionization potential, from \civ\ (C$^{3+}$) to \aliii\ and down to \mgii. While \civ\ blueshifts can exceed several thousand \kms, \mgii\ shows only subtle but measurable shifts, detectable mainly in extreme Pop. A quasars. These trends are consistent with an ionization-stratified broad-line region, where higher ionization lines originate closer to the black hole in regions more directly exposed to radiation pressure (Figure \ref{fig:disks}). 

Photoionization simulations with realistic high-accretor SEDs demonstrate that outflows can be launched over a wide range of radii if the column density remains low ($N_\mathrm{c} \approx 10^{21}$ cm$^{-2}$), whereas dense gas ($N_\mathrm{c} \approx 10^{24}$ cm$^{-2}$) remains gravitationally bound. These models reproduce both the hierarchy of shift amplitudes (\civonly\ $>$ \aliiionly\ $>$ \mgiionly) and the reverberation-mapped radii of \civ\ and \mgii\ \citep{ homayounietal20, khadkaetal21, caoetal22, shenetal24}, providing a physically consistent picture that links SED shape, BLR stratification, and wind dynamics. %In this framework, extreme Population A quasars represent the most efficient accretors, where radiation pressure dominates not only the dynamics of high-ionization gas but also leaves a measurable imprint on intermediate- and low-ionization lines. 

\acknowledgements

S. P. is supported by the international Gemini Observatory, a program of NSF NOIRLab, which is managed by the Association of Universities for Research in Astronomy (AURA) under a cooperative agreement with the U.S. National Science Foundation, on behalf of the Gemini partnership of Argentina, Brazil, Canada, Chile, the Republic of Korea, and the United States of America. A. del Olmo and A. Deconto-Machado acknowledge financial support from the grant PID2022-140871NB-C21 funded by MCIN/AEI/10.13039/501100011033 and by ‘ERDF A way of making Europe’, and through the Center of Excelence Severo Ochoa grant CEX2021- 515001131-S of the IAA funded by MCIN/AEI/10.13039/501100011033. 

\clearpage

%%%%%%%%%%%%%%%%%%%%%%%%%%%%%%%%%%%%%%%%%%%%%%%%%%%%%%%%%%%%%%%%%%%%%%%%%%%%%
%                       H A P P Y E N D                                      
% Your LaTeX source text must be ended by the line:                          
%%%%%%%%%%%%%%%%%%%%%%%%%%%%%%%%%%%%%%%%%%%%%%%%%%%%%%%%%%%%%%%%%%%%%%%%%%%%%
\end{document}